\documentstyle[12pt]{article}
\def\ln{{\rm ln}}

\def\a{\begin{eqnarray}}
\def\b{\end{eqnarray}}
\def\0{\nonumber}
\def\ba{\begin{array}}
\def\ea{\end{array}}
\def\noal{\noalign{\vskip10pt}}

\def\q{{\bar{\cal Q}}}

\def\al{{\alpha}}
\def\lm{{\lambda}}

\def\cm{{\cal M}}

\renewcommand{\theequation}{\thesection.\arabic{equation}}

\newlength{\extraspace}
\newlength{\extraspaces}
\newcounter{dummy}
\newcommand{\ai}{
\addtocounter{equation}{1}
\setcounter{dummy}{\value{equation}}
\setcounter{equation}{0}
\renewcommand{\theequation}{\thesection.\arabic{dummy}\alph{equation}}
\begin{eqnarray}
\addtolength{\abovedisplayskip}{\extraspaces}
\addtolength{\belowdisplayskip}{\extraspaces}
\addtolength{\abovedisplayshortskip}{\extraspace}
\addtolength{\belowdisplayshortskip}{\extraspace}}
\newcommand{\bj}{
\end{eqnarray}
\setcounter{equation}{\value{dummy}}
\renewcommand{\theequation}{\thesection.\arabic{equation}}}
\def\d{{\partial}}

\newcommand{\ddlm}[1]{{\partial \over \partial \lm_{#1}}}

\begin{document}
\begin{flushright}
SISSA-ISAS 54/94/EP\\
BONN-HE-06/94\\
hep-th/9405004
\end{flushright}
\vskip0.5cm
\centerline{\LARGE\bf Two--matrix model and $c=1$ string theory}
\vskip0.3cm
\centerline{\large  L.Bonora}
\centerline{International School for Advanced Studies (SISSA/ISAS)}
\centerline{Via Beirut 2, 34014 Trieste, Italy}
\centerline{INFN, Sezione di Trieste.  }
\vskip0.5cm
\centerline{\large C.S.Xiong}
\centerline{Physikalisches Institut der Universit\"at Bonn}
\centerline{Nussallee 12, 53115 Bonn, Germany}
\vskip3cm
\abstract{We show that the most general two--matrix model with bilinear
coupling underlies $c=1$ string theory. More precisely we prove that
$W_{1+\infty}$ constraints, a subset of the correlation functions and the
integrable hierarchy characterizing such two--matrix model, correspond
exactly to the $W_{1+\infty}$ constraints, to the discrete tachyon
correlation functions and to the integrable hierarchy of the $c=1$ string
theory.}

\vskip3cm

\section{Introduction}

$c=1$ string theory is a well--studied subject. The spectrum of the states
is well--known, the correlation functions of the so--called discrete
tachyons have been calculated at least in low genus, the underlying $W$
constraints have been written down in explicit form (for reviews on the
subject, see for example \cite{review}). Recently the attention
of several people has shifted to the problem of identifying the integrable
hierarchy underlying $c=1$ string theory and the topological field theory  it
may be related to. The hierarchy proposed \cite{Taka},\cite{EK} is
the dispersionless Toda lattice hierarchy. A Landau--Ginzburg
potential has also been identified
\cite{GM},\cite{Oz},\cite{Taka},\cite{EK} for this theory.

The purpose of this letter is to show that the two--matrix model with
bilinear coupling \cite{BX1} provide a unified framework for all these
scattered elements of $c=1$ the theory. In fact in section 2, after recalling
the main features of two--matrix models,  we
show that the  $W_{1+\infty}$ constraints proposed in order to interpret the
$c=1$ string theory are nothing but the $W_{1+\infty}$ constraints of
two--matrix models. Using the latter in section 3 we show how to calculate
correlation functions of discrete tachyons and verify that they coincide
at least in genus 0 with the ones in the literature. Section 4 is devoted
to the comparison between the hierarchies: it was found in \cite{BX1} that
two--matrix models are characterized by the Toda lattice hierarchy;
we show that in genus 0 this is exactly the integrable structure suggested
in \cite{Taka} and in \cite{EK}. We show in particular that
some ad hoc assumptions made in the latter references are a natural
consequence of the two--matrix model structure.

\section{The two--matrix model and its $W_{1+\infty}$ constraints}

Two--matrix models are defined by the partition function
\a
Z_N(t,c)=\int dM_1dM_2 e^{TrU}\0
\b
where $M_1$ and $M_2$ are Hermitian $N\times N$ matrices and
\a
U=V_1 + V_2 + g M_1 M_2\0
\b
with potentials
\a
V_{\al}=\sum_{r=1}^{p_{\al}}t_{\al,r}M_{\al}^r\,\qquad \al=1,2.\label{V}
\b
The $p_{\al}$'s are positive integers, characterizing the particular model.
In this paper we consider $p_1$ and $p_2$ as arbitrarily large numbers and
denote them simply as $\infty$. In the following we refer to this model as
{\it the} two--matrix model.

We showed in \cite{BX1} that one can map this functional integral problem
into a linear integrable system together with definite coupling constraints.
For the sake of clarity and with the purpose of introducing formulas and
notation we will need later, we review this part once again.
The ordinary procedure to calculate the partition function consists of
three steps \cite{IZ2},\cite{M}:
$(i)$ one integrates out the angular parts so that only the
integrations over the eigenvalues are left;
$(ii)$ one introduces the orthogonal polynomials
\a
\xi_n(\lambda_1)=\lambda_1^n+\hbox{lower powers},\qquad\qquad
\eta_n(\lambda_2)=\lambda_2^n+\hbox{lower powers}\0
\b
which satisfy the orthogonality relations
\a
\int  d\lambda_1d\lambda_2\xi_n(\lambda_1)
e^{V_1(\lm_1)+V_2(\lm_2)+g\lm_1\lm_2}
\eta_m(\lambda_2)=h_n(t,g)\delta_{nm}\label{orth1}
\b
$(iii)$, using the orthogonality relation (\ref{orth1}) and the properties
of the Vandermonde determinants, one can easily
calculate the partition function
\a
Z_N(t,g)={\rm const}~N!\prod_{i=0}^{N-1}h_i\label{parti1}
\b
Knowing the partition function means knowing
the coefficients $h_n(t,g)$'s.

The information concerning the latter
can be encoded in a suitable linear system plus some coupling conditions,
together with the reconstruction formulas for $Z_N$.
But before we pass to that we need some convenient notations.
For any matrix $M$, we define
\a
\bigl(\cm\bigl)_{ij}= M_{ij}{{h_j}\over{h_i}},\qquad
{\bar M}_{ij}=M_{ji},\qquad
M_l(j)\equiv M_{j,j-l}.\0
\b
As usual we introduce the natural gradation
\a
deg[E_{ij}] = j -i\0
\b
and, for any given matrix $M$, if all its non--zero elements
have degrees in the interval $[a,b]$, then we will simply
write: $M\in [a,b]$. Moreover $M_+$ will denote the upper triangular
part of $M$ (including the main diagonal), while $M_-=M-M_+$. We will write
\a
{\rm Tr} (M)= \sum_{i=0}^{N-1} M_{ii}\0
\b

Let us come now to the above mentioned linear system and coupling conditions.
First it is convenient to pass from the basis of orthogonal polynomials
to the basis of orthogonal functions
\a
\Psi_n(\lambda_1)=e^{V_1(\lambda_1)}\xi_n(\lambda_1),
\qquad
\Phi_n(\lambda_2)=e^{V_2(\lambda_2)}\eta_n(\lambda_2).\0
\b
The orthogonality relation (\ref{orth1}) becomes
\a
\int d\lm_1 d\lm_2\Psi_n(\lambda_1)e^{g\lm_1\lm_2}
\Phi_m(\lambda_2)=\delta_{nm}h_n(t,g).\label{orth2}
\b
As usual we will denote the semi--infinite column vectors with components
$\Psi_0,\Psi_1,\Psi_2,\ldots,$ and  $\Phi_0,\Phi_1,\Phi_2,\ldots,$
by $\Psi$ and $\Phi$, respectively.

Next we introduce the following $Q$--type matrices
\a
\int d\lm_1 d\lm_2\Psi_n(\lambda_1)
\lm_{\al}e^{g\lm_1\lm_2}
\Phi_m(\lambda_2)\equiv Q_{nm}(\al)h_m=\q_{mn}(\al)h_n,\quad
\al=1,2.\label{Qalpha}
\b
Both $Q(1)$ and $\q(2)$ are
Jacobi matrices: their pure upper triangular
part is $I_+=\sum_i E_{i,i+1}$.

Beside the above $Q$ matrices, we will need two $P$--type matrices, defined by
\a
&&\int d\lm_1 d\lm_2\Bigl(\ddlm 1 \Psi_n(\lambda_1)\Bigl)
e^{g\lm_1\lm_2}\Phi_m(\lambda_2)\equiv P_{nm}(1)h_m\\
&&\int  d\lambda_1d\lambda_2\Psi_n(\lambda_1)e^{g\lm_1\lm_2}
\Bigl(\ddlm 2 \Phi_m(\lambda_2)\Bigl)\equiv P_{mn}(2)h_n
\b

The following relations hold
\a
P(1)+g Q(2)=0,\qquad\quad
gQ(1)+\bar{\cal P}(2)=0,\label{coupling}
\b
It is just these coupling conditions that lead to the famous
$W_{1+\infty}$--constraints on the partition function.
{}From them it follows at once that
\a
Q(1)\in[-\infty, 1],\qquad Q(2) \in [-1 , \infty]\0
\b

The derivation of the  linear systems associated to the two--matrix model
is very simple.  We take the derivatives of eqs.(\ref{orth2})
with respect to the time parameters $t_{\al,r}$, and use
eqs.(\ref{Qalpha}).  We get in this way the time evolution of $\Psi$
and the {\it first discrete linear system}:
\a
\left\{\ba{ll}
Q(1)\Psi(\lambda_1)=\lambda_1\Psi(\lambda_1),& \\\noal
{\partial\over{\partial t_{1,k}}}\Psi(\lambda_1)=Q^k_+(1)
\Psi(\lambda_1),&\\\noal
{\partial\over{\partial t_{2,k}}}\Psi(\lambda_1)=-Q^k_-(2)
\Psi(\lambda_1),&\\\noal
{\partial\over{\partial\lm}}\Psi(\lambda_1)=P(1)\Psi(\lm_1).&
\ea\right.\label{DLS1}
\b
The corresponding consistency conditions are
\ai
&&[Q(1), ~~P(1)]=1\label{CC11}\\
&&{\partial\over{\partial t_{\al,k}}}Q(1)=[Q(1),~~Q^k_-(\al)],
\label{CC12}\\
&&{\partial\over{\partial t_{1,k}}}P(1)=[Q^k_+(1), P(1)],\quad\quad
{\partial\over{\partial t_{2,k}}}P(1)=[P(1),~~Q^k_-(2)]
\label{CC13}
\bj
where $\alpha =1,2$.

By studying the evolution of $\Phi$ we get the {\it second discrete linear
system}
\a
\left\{\ba{ll}
\bar {\cal Q}(2)\Phi(\lambda_2)=\lambda_2\Phi(\lambda_2),& \\\noal
{\partial\over{\partial t_{2,k}}}\Phi(\lambda_2)=\bar {\cal Q}^k_+(2)
\Phi(\lambda_2),&\\\noal
{\partial\over{\partial t_{1,k}}}\Phi(\lambda_2)=-\bar {\cal Q}^k_-(1)
\Phi(\lambda_2),&\\\noal
{\partial\over{\partial\lm}}\Phi(\lambda_2)=P(2)\Psi(\lm_2).&
\ea\right.\label{DLS2}
\b
The corresponding consistency conditions are
\ai
&&[\bar {\cal Q}(2), ~~P(2)]=1\label{CC21}\\
&&{\partial\over{\partial t_{\al,k}}}\bar {\cal
Q}(2)=[\bar {\cal Q}(2),~~\bar {\cal Q}^k_-(\al)], \label{CC22}\\
&&{\partial\over{\partial t_{1,k}}}P(2)=[P(2),~~\bar {\cal
Q}^k_-(1)],\quad\quad
{\partial\over{\partial t_{2,k}}}P(2)=[\bar {\cal Q}^k_-(2), P(2)]
\label{CC23}
\bj
where $\alpha =1,2$. Eqs.(\ref{CC12}),(\ref{CC13}),(\ref{CC22}),(\ref{CC23})
define the Toda lattice hierarchy.

The third element we need is the link between the
quantities that appear in the linear systems and in the coupling conditions
with the original partition function. We have
\a
{\d \over \d_{\al, r}} \ln Z_N(t,g) = {\rm Tr} \Big(Q^r(\al)\Big), \quad\quad
\al = 1,2 \label{ddZ}
\b
It is evident that, by using the flow equations above we can express all
the derivatives of $Z_N$ in terms of the elements of the $Q$ matrices. For
example
\a
{\d^2\over{\d t_{1,1}\d t_{\al,r}}}
\ln Z_N(t,g)=\Bigl(Q^r(\al)\Bigl)_{N,N-1},\quad \al =1,2\label{parti3}
\b
Knowing all the derivatives with respect to the coupling parameters
we can reconstruct the partition function up to an overall integration
constant.

We also remark that, since the RHS of the above equations is always defined,
they give us a definition of $Z_N$ even in subsets of the parameter space where
the path--integral is ill--defined.

Another consequence of eq.(\ref{ddZ}) and of the definitions of $Q(\al)$ and
$\Psi$, is that we can write
\a
\ln \psi_N = \sum_{r=1}^\infty t_r \lambda_1^r +N{} \ln \lambda_1 -
\sum_{r=1}^\infty {1\over {r \lambda_1^r}} {\d\over {\d t_r}} \ln
Z_N\label{Psi} \b
from which we have
\a
P(1) = \sum_{r=1}^\infty rt_r Q(1)^{r-1} + N Q(1)^{-1} +
\sum_{r=1}^\infty Q(1)^{-r-1} {{\d \ln Z_N}\over{\d t_r}}\label{P}
\b
Similar formulas hold for $\ln \Phi_n$ and $P(2)$.

We will be using the following coordinatization of the Jacobi matrices
\a
Q(1)=I_++\sum_i \sum_{l=0}^\infty a_l(i)E_{i,i-l}, \qquad\qquad\qquad
\q(2)=I_++\sum_i \sum_{l=0}^\infty b_l(i)E_{i,i-l}\label{jacobi}
\b
One can immediately see that, for example,
\a
\Bigl(Q_+(1)\Bigl)_{ij}=\delta_{j,i+1}+a_0(i)\delta_{i,j},\qquad
\Bigl(Q_-(2)\Bigl)_{ij}=R_i\delta_{j,i-1}\label{jacobi1}
\b
where $R_{i+1} \equiv h_{i+1}/h_i$.
As a consequence of this coordinatization, eq.(\ref{parti3}) gives in
particular the important relation
\a
{\d^2\over{\d t_{1,1}\d t_{2,1}}}\ln Z_N(t, g) = R_{N-1} \label{ZR}
\b
Finally let us quote from \cite{BX1} the equation
\a
{{\d^2 }\over{\d t_{1,1} \d t_{2,1}}} \ln R_j = R_{j+1}-2R_j
+R_{j-1}\label{Todaeq}
\b
which justifies the name given to the hierarchy.

\subsection{$W_{1+\infty}$ constraints}

The $W_{1+\infty}$ constraints (or simply $W$--constraints)
on the partition function for our two--matrix
model were obtained in \cite{IM},\cite{BX1}, by putting together both coupling
conditions and consistency conditions. In other words
the $W_{1+\infty}$ constraints contain all the available information.
They take the form
\a
W^{[r]}_n Z_N(t,g)=0, \quad\quad\quad
\tilde W^{[r]}_n Z_N(t,g)=0\quad r\geq0;~~n\geq-r,\label{Wc}
\b
where
\ai
W^{[r]}_n&\equiv& (-g)^n{\cal L}^{[r]}_n(1)-{\cal
L}^{[r+n]}_{-n}(2)\label{Wa}\\
\tilde W^{[r]}_n&\equiv& (-g)^n{\cal L}^{[r]}_n(2)-{\cal L}^{[r+n]}_{-n}(1)
\label{Wb}
\bj

The generators ${\cal L}^{[r]}_n(1)$ are differential operators involving
$N$ and $t_{1,k}$, while ${\cal L}^{[r]}_n(2)$ have the same form
with $t_{1,k}$ replaced by $t_{2,k}$. One of the remarkable
aspects of (\ref{Wc}) is that the dependence on the coupling $g$ is nicely
factorized.
The $W_{1+\infty}$ algebra satisfied by the ${\cal L}^{[r]}_n(1)$ has been
written down in ref.\cite{BX1},\cite{BX2}.
In general we have
\a
[{\cal L}^{[r]}_n(1), {\cal L}^{[s]}_m(1)]=(sn-rm)
 {\cal L}^{[r+s-1]}_{n+m}(1)+\ldots,\label{LLgen}
\b
for $r,s\geq 1;~n\geq-r,m\geq-s$. Here dots denote lower than $r+s-1$ rank
operators.
The algebra of the ${\cal L}^{[r]}_n(2)$ is just a copy of the above one,
and the algebra satisfied by the $W^{[r]}_n$ and by the $\tilde W^{[r]}_n$
is isomorphic to both.

There is a sometimes simpler way to write the above generators. It consists
in introducing the U(1) current
\a
J(z) = \sum_{r=1}^\infty r t_r z^{r-1} + {N \over z} + \sum_{r=1}^\infty
z^{-r-1} {\d \over {\d t_r}}\label{J}
\b
and defining the density
\a
{\cal L}^{[n]}(z) = {1\over{n+1}} :\Bigl( -\d + J(z)\Bigl)^{n+1}: \cdot 1 \0
\b
Then ${\cal L}_k^{[n]}$ can be recovered as
\a
{\cal L}_k^{[n]} = Res_{z=0} ({\cal L}^{[n]}(z) z^{n+k})\0
\b
The above definition holds for both the 1 and 2 sector.
One can also consider the fermionized version of the above formulas. This
leads us to the $W_{1+\infty}$ constraints suggested in ref.\cite{DMP}
for the $c=1$ string theory.
One easily realizes, by using either the bosonic or the fermionic
representation, that the latter are a subset of the constraints (\ref{Wc}).
In fact they coincide with the cases $n = -r$ and $g=-1$ and can be written
explicitly in the form
\a
&&{\d \over {\d t_{1,r}}}Z_N = {\cal L}_{-r}^{[r]}(2) Z_N\label{DMPa}\\
&&{\d \over {\d t_{2,r}}}Z_N = {\cal L}_{-r}^{[r]}(1) Z_N\label{DMPb}
\b
{\it This is our first link between two--matrix model and $c=1$ string
theory}.

\section{Correlation functions of discrete tachyons}

In this section, from the $W$ constraints (\ref{Wc}) we calculate
a subset of the correlations functions of two--matrix model in a
very simple {\it small phase space} and identify them with the
correlation functions of the discrete tachyons of the $c=1$ string theory.
{}From now we set $g = -1$.

To start with let us introduce some simplified notations:
\a
t_{1,k} \equiv t_k ,&& \quad\quad\quad t_{2,k} \equiv s_k\0\\
{\rm Tr}M_1^k \equiv \tau_k, &&\quad\quad\quad
{\rm Tr}M_2^k \equiv \sigma_k\0
\b
The correlation functions of the two--matrix model are defined by
\a
\ll \tau_{k_1}\ldots \tau_{k_n} \sigma_{l_1}\ldots \sigma_{l_m}\gg =
{\partial\over{\partial t_{k_1}}}
 \ldots {\partial\over{\partial t_{k_n}}} {\partial\over{\partial s_{l_1}}}
\ldots {\partial\over{\partial s_{l_m}}} \ln Z_N(t,s)\0
\b

Let us write some of the $W$--constraints in this new language. $W^{[1]}_{-1}
Z_N=0$ and $\tilde W^{[1]}_{-1} Z_N=0$ become, respectively
\ai
&&\sum_{k=2}^\infty k t_k \ll \tau_{k-1} \gg  +N t_1 - \ll \sigma_1\gg =0
\label{w1-1a}\\
&&\sum_{k=2}^\infty k s_k \ll \sigma_{k-1} \gg  +N s_1 - \ll \tau_1\gg =0
\label{w1-1b}
\bj
Instead $W^{[1]}_{0} Z_N=0$ and $\tilde W^{[1]}_{0} Z_N=0$ give rise to
the same equation
\a
\sum_{k=1}^\infty k t_k \ll \tau_k \gg =
\sum_{k=1}^\infty k s_k \ll \sigma_k \gg \label{w10}
\b
and so on.

The $W$ constraints considered so far are exact, they contain
contributions from all the genera. For simplicity we limit ourselves in
this section to the genus 0 contribution. The $W$ constraints assume,
in this case, a simplified expression. To find it one can follow the
homogeneity analysis of \cite{BX2}. Equivalently one can define
$x = N/{\beta}$, consider the rescalings
\a
t_k \rightarrow {t_k\over{\beta}}, \qquad s_k \rightarrow {s_k\over \beta},
\qquad c \rightarrow {c \over \beta}, \qquad{\cal F}^{(0)} \rightarrow
\beta^{-2} {\cal F}^{(0)}\0
\b
and keep the leading term in $\beta$ in (\ref{Wc}). Here $\beta$ is an
arbitrary positive number and ${\cal F}^{(0)}$ is the genus 0
part of $\ln Z_N$. Once this is done one quickly realizes that
the genus 0 version
of (\ref{w1-1a}), (\ref{w1-1b}) and (\ref{w10}) remain (formally) the same,
except for the fact that $N$ is replaced by $x$ and now $\ll \cdot \gg$
denotes only the genus 0 contribution, not the complete correlation function.
The latter simplifying convention will be followed until the end of this
section. However the higher $W$ constraints in general change form when
reduced to genus 0. For example, $W_{-2}^{[2]}Z_N=0$ becomes
\a
&& \sum_{l_1, l_2=1}^\infty l_1 l_2 l t_{l_1}t_{l_2} \ll \tau_{l_1+l_2-2}\gg
+\sum_{l=4}^\infty l t_l \sum_{k=1}^{l-3}
\ll \tau_k \gg\ll \tau_{l-k-2}\gg\0\\
&&~~~~~~~~~~~~~~+2x \sum_{l=3}^\infty l t_l \ll \tau_{l-2} \gg + 2x^2  t_2 +
  x t_1^2 ~=~  \ll \sigma_2 \gg
\label{w2-2}
\b

Next we specify the small phase space in which we want to compute our
correlation functions. This is the simplest possible one: $t_k=0=s_k$,
$\forall k$. We will denote the correlation functions calculated in
such restricted parameter space as $<\cdot>$, instead of $\ll\cdot\gg$.

Now we can set out to calculate the correlation functions. First of all
let us differentiate both sides of (\ref{w10}) with respect to
$t_{k_1}, \ldots, t_{k_n}$ and $t_{l_1}, \dots , t_{l_m}$. We get
\a
(k_1 + \ldots + k_n - l_1 - \ldots - l_m) <\tau_{k_1} \dots \tau_{k_n}
\sigma_{l_1} \dots \sigma_{l_m}> =0\0
\b
which means that the correlation functions are nonvanishing only when
\a
k_1 + \ldots + k_n - l_1 - \ldots - l_m=0\label{deltacc}
\b
We remark that this can be interpreted as a charge conservation.
Next from (\ref{w1-1a}, \ref{w1-1b}, \ref{w2-2}) we get immediately
\a
<\sigma_1\tau_1> = x, \qquad <\sigma_2\tau_2> = 2 x^2 \0
\b
In general from (\ref{DMPa}, \ref{DMPb}) we get
\a
<\tau_l, \sigma_l > = l x^l \0
\b
Next one differentiate (\ref{w1-1a}) w.r.t. $t_l$ and $s_{l-1}$, and
(\ref{w2-2}) w.r.t. $t_l, s_{l-2}$, and obtain
\a
&&<\sigma_1 \sigma_{l-1} \tau_l> = l <\tau_{l-1} \sigma_{l-1}>\0\\
&&<\sigma_2 \sigma_{l-2} \tau_l > = 2 x l <\sigma_{l-2} \tau_{l-2}>\0
\b
and so on. From this it is easy to conclude that
\a
<\sigma_k \sigma_{l-k} \tau_l> = kl(l-k) x^{l-1}\label{3p}
\b
Proceeding the same way it is not difficult to arrive at
\a
<\tau_k \sigma_{l_1} \sigma_{l_2}\ldots \sigma_{l_m}> =
k l_1l_2\ldots \l_m (k-1)(k-2) \ldots (k-m+2) x^{k-m+1}\label{cf1}
\b
where $k= l_1 + l_2 +\ldots + l_m$, $m\geq3$. If we rescale $t_k \rightarrow
kt_k$
and $s_k \rightarrow k s_k$, the new correlation functions (\ref{cf1})
become
\a
<\tau_k \sigma_{l_1} \sigma_{l_2}\ldots \sigma_{l_m}> =
(k-1)(k-2) \ldots (k-m+2) x^{k-m+1}\label{cf2}
\b
These expressions are pretty familiar to those who are acquainted with the
literature on the $c=1$ string theory. If we set
\a
{\cal T}_k \equiv \tau_k, \qquad {\cal T}_{-k} \equiv \sigma_k\label{redef}
\b
and interpret $x$ as the cosmological constant,
the correlation functions (\ref{cf2}) and the like are nothing but the
correlation functions of the discrete tachyons ${\cal T}_k$ calculated
in terms of the cosmological constant alone, \cite{many}.

{\it This is our second link between two--matrix model and $c=1$ string
theory.}

We remark that one can obtain many more results beside the genus 0
correlation functions in the small phase space given above. One can
explicitly calculate correlation functions in higher genus and in a larger
space of parameters. However, since in this letter, our concern
is to motivate the connection between two--matrix models and $c=1$ string
theory, we postpone these calculations to a future publication.

\section{The dispersionless Toda lattice hierarchy}

We have already pointed out that the integrable hierarchy that appears in
our two--matrix model is the Toda lattice hierarchy (see also \cite{BX2}).
In two recent papers, \cite{Taka} and \cite{EK}, it has been suggested
that the dispersionless Toda hierarchy underlies the Landau--Ginzburg
formulation of the $c=1$ string theory. To those familiar with integrable
hierarchies it is almost evident that our integrable hierarchies (\ref{CC12},
\ref{CC13},\ref{CC22},\ref{CC23}) coincide in the dispersionless limit
with the dispersionless Toda lattice hierarchy of \cite{Taka} and \cite{EK}.
It is
also  apparent that the constraints introduced {\it ad hoc} in the latter
references are nothing but our coupling conditions (\ref{coupling}), which
are natural relations in the framework of two--matrix models.

However, for most these are rather technical issues. Therefore we
will spend this section to explicitly show the identification we
just claimed. As our first step we rewrite the formulas of section 2 in the
dispersionless limit. This can be achieved with a continuum
limit. We showed in a series of papers (see, for example \cite{BX3},
\cite{BX1}), that there is actually no need to take this continuum limit, we
can obtain the same results more neatly with the discrete approach,
especially when all the genera are involved. However here we have to
make a comparison with a continuum formulation. Therefore in this paper
we will shift to it.

The continuum dispersionless limit is obtained by promoting the matrix
index $n$ to a continuum variable $x$ and by introducing the conjugate
variable $\zeta$
\a
\{ \zeta , x \} = \zeta\label{FP}
\b
We recall that the variables $x$ and $\zeta$ can be traced back to the
discrete matrices $\rho = \sum_n n E_{n,n}$ and $I_+$ \cite{BX3},
respectively, and the
Poisson bracket (\ref{FP}) is nothing but the continuum version of the
commutator
\a
\relax [ I_+, \rho] = I_+\0
\b

In this limit the `fields' $a_l(n)$ and $b_l(n)$ on the lattice (see
eq.(\ref{jacobi})) are
mapped into fields which are function of $x$ (beside the coupling constants).
Therefore we are going to have the following replacements
\a
&&Q(1) \rightarrow L = \zeta + \sum_{r=0}^\infty a_r
\zeta^{-r}\label{trans1}\\
&&\bar {\cal Q}(2) \rightarrow {\tilde L} = \zeta + \sum_{r=0}^\infty b_r
\zeta^{-r}\label{trans2}
\b
Similarly
\a
&&P(1) \rightarrow M= \sum_{r=1}^\infty rt_r L^{r-1} + x L^{-1} +
\sum_{r=1}^\infty {{\d {\cal F}^{(0)}}\over {\d t_r}}L^{-r-1}\label{trans3}\\
&&P(2) \rightarrow {\tilde M} = \sum_{r=1}^\infty rs_r {\tilde L}^{r-1} + x
{\tilde L}^{-1} +
\sum_{r=1}^\infty {{\d {\cal F}^{(0)}}\over {\d s_r}}{\tilde
L}^{-r-1}\label{trans4}
\b
Moreover
\a
\ln \Psi_N \rightarrow \ln \Psi = \sum_{r=1}^\infty t_r \lambda_1^r + x
\ln \lambda_1 - \sum_{r=1}^\infty {1\over {r \lambda_1^r}} {\d\over {\d t_r}}
{\cal F}^{(0)}\label{logpsi}
\b

Now we have to introduce the continuous analog of the operation that
maps a matrix $Q$ to the matrix $\bar {\cal Q}$, which we defined at the
beginning. This is the operation $\sigma$ which maps
\a
\sigma(\zeta) = {R\over \zeta} \label{sigma}
\b
and leave unaltered all the other quantities. Here $R$ is the continuous
limit of the `field' $R_n$ which we have defined after eq.(\ref{jacobi1}).

The form the coupling constraints (\ref{coupling}) take
in the dispersionless limit (remember that $g =-1$), is
\a
M = \sigma(\widetilde L), \qquad L = \sigma({\widetilde M})\label{newcoupl}
\b
respectively.

We can now write down the dispersionless versions of the hierarchies
(\ref{CC12}, \ref{CC13}, \ref{CC22}, \ref{CC23}):
\a
&&{{\d L}\over {\d t_r}} = \{ L^r_+ ,L\}, \quad\quad {{\d L}\over {\d s_r}}
=\{L, M^r_-\} \label{dispCC1}\\
&&{{\d M}\over {\d t_r}} = \{ L^r_+ ,M\}, \quad\quad {{\d M}\over {\d s_r}}
=\{M, M^r_-\} \label{dispCC2}
\b
Here the subscript + means the part of an expression contaning non--negative
powers of  $\zeta$, while the -- indicates the complementary part.
Isomorphic hierarchies can be obtained by applying the $\sigma$ operator
to these equations.

It is now easy to make a comparison with refs.\cite{Taka} and \cite{EK}
and verify that the hierarchies are the same. To be more precise
the correspondence of our paper with, for example, ref.\cite{EK} is (our
notations are on the left hand side)
\a\left.\begin{array}{ll}
L \leftrightarrow L, \qquad\qquad &M\leftrightarrow ML^{-1}\\
x  \leftrightarrow \mu , \qquad\qquad &\zeta  \leftrightarrow x \\
 \widetilde M  \leftrightarrow \bar M \bar L , \qquad\qquad
&\sigma(\widetilde L)  \leftrightarrow {\bar L}^{-1}\end{array}\right.\0
\b
Moreover the constraints (23),(24) of \cite{EK} are nothing but
(\ref{newcoupl}) above, and so on. It is apparent that the definitions and
constraints introduced {\it ad hoc} in refs.\cite{Taka} and \cite{EK}
are completely natural in the framework of the two--matrix model.

{\it This is the third element in common between the two--matrix model
and the $c=1$ string theory}.

It has been suggested that the $c=1$ string theory can be given a topological
Landau-Ginzburg interpretation,
\cite{GM},\cite{Oz},\cite{EK},\cite{Taka}. In our language the
Landau--Ginzburg potential proposed by these authors coincides with $M$. In
fact let us define
\a
\phi_i = {{\d M}\over {\d t_i}}, \quad \phi_0 = {{\d M}\over {\d x}}, \quad
\phi_{-j} = {{\d M}\over {\d s_j}}\label{defphi}
\b
We notice that if $t_i=0=s_j$ $\forall i,j$, then
\a
M= {x\over \zeta}, \qquad L=\zeta=\lambda_1\0
\b
and
\a
\phi_i = i \zeta^{i-1}, \qquad \phi_0 = {1\over \zeta}, \qquad
\phi_{-j} = j x^j \zeta^{-j-1} \0
\b
If we define
\a
<\phi_i \phi_j \phi_k> = - Res_{\zeta=0} {{\phi_i \phi_j \phi_k}\over {M'}}\0
\b
where $M' = {{\d M}\over {\d \zeta}}$, we find immediately that
these three point function coincides with (\ref{3p}). After a trivial
redefinition the $\phi_i$'s are the Landau-Ginzburg representatives of the
${\cal T}_i$ if $M$ is assumed to be the corresponding potential.

\section{Discussion}

One could add other elements to the correspondence between the two--matrix
model and the $c=1$ string theory. A short example is the following.
Eq.(\ref{Todaeq}), in the dispersionless limit, reads
\a
{{\d}\over {\d t_1 \d s_1}} \ln R = {{\d ^2 R }\over {\d^2 x}}\label{disptoda}
\b
On the other hand, from eq.(\ref{ZR}) we have
\a
R = {{\d^2 }\over {\d t_1 \d s_1}} {\cal F}^{(0)}\0
\b
But the RHS of this equations is nothing but the correlation function
calculated in the previous section, namely $<\tau_1\sigma_1>= x$. Putting
all this together we obtain
\a
u^{(0)} \equiv {{\d^2}\over {\d x^2 }} {\cal F}^{(0)} = \ln R = \ln
x\label{chem}
\b
This is the expected behaviour of the chemical potential in terms of
the cosmological constant in the $c=1$ string theory in genus 0.

Even though we have not discussed
the subject thoroughly (in particular for lack of space we postpone
a discussion of the discrete states),
we think we have given enough evidence that a two--matrix model
underlies the $c=1$ string theory. This claim may sound a priori surprising,
since we are familiar with the idea that the $c=1$ string theory is described
by a model of one time--dependent matrix. However it is more
plausible than it seems at first sight, if one thinks that the connection
between two--matrix model and $c=1$ string theory is at the level
of topological degrees of freedom. From this point of view let us recall
that in \cite{BX2} we showed
two matrix--models contain all n--KdV hierarchies and the relevant
$A_n$ series of topological field theories. It has
been suggested recently \cite{AK} that they might contain also the $D$ series
of topological field theories and, perhaps, other series. We do think
that the two--matrix models are a perfect framework for a large family
of topological field theories and will describe the topological field
theory content of it in a forthcoming work.

\end{document}